\documentclass[aps,superscriptaddress,prl,reprint,
 showpacs ,amsmath,amssymb,twocolumn]{revtex4-1}

\usepackage{graphicx}

\usepackage{amsmath}
\newcommand{\rest}{\mathrm r}
\newcommand{\thr}{{\mathrm {thr}}}
\newcommand{\ext}{{\mathrm {ext}}}
\newcommand{\fb}{{\mathrm {fb}}}

\newcommand{\gs}{{g_{\mathrm s}}}
\newcommand{\taum}{{\tau_{\mathrm m}}}
\newcommand{\taud}{{\tau_{\mathrm d}}}
\newcommand{\tausone}{{\tau_{\mathrm s1}}}
\newcommand{\taustwo}{{\tau_{\mathrm s2}}}
\newcommand{\tp}{{t^\prime}}
\newcommand{\tpp}{{t^{\prime\prime}}}

\newcommand{\xp}{{x^\prime}}
\newcommand{\Laplace}{\mathrm L}

\newcommand{\eq}{\mathrm {eq}}
\newcommand{\xav}{x_{\mathrm {ave}}}
\newcommand{\xaveq}{x_{\mathrm {ave,eq}}}

\begin{document}

\title{Renewal Approach to the Analysis of the Asynchronous State for 
Coupled Noisy Oscillators}

\author{Farzad Farkhooi} \affiliation{Neuroinformatics \& Theoretical
Neuroscience, Freie Universit{\"a}t Berlin and BCCN-Berlin, Germany}

\author{Carl van Vreeswijk} \affiliation{Centre de Neurophysique Physiologie et
Pathologie, ParisDescartes University and CNRS UMR 8119, Paris, France}

\date{\today}

\begin{abstract}

We develop a framework in which the activity of nonlinear pulse-coupled
oscillators is posed within the renewal theory. In this approach, the evolution
of inter-event density allows for a self-consistent calculation that determines
the asynchronous state and its stability. This framework, can readily be
extended to the analysis of systems with more state variables. To exhibit this,
we study a nonlinear pulse-coupled system, where couplings are dynamic and
activity dependent. We investigate stability of this system and we show it
undergoes a super-critical Hopf bifurcation to collective synchronization.
\end{abstract}


\maketitle

The collective dynamics of pulse-coupled networks of nonlinear 
oscillators has been studied extensively \cite{kuramoto_chemical_2003,
*winfree_geometry_2001}. In the presence of noise, the approach has been to
analyze the Fokker-Planck equation of the state variables of the
oscillators. This has been highly successful for systems with units described
by a single variable. However, it has proven to be difficult to extend this
approach to systems with several state variables. This is because in such
systems the Fokker-Planck equation may have a highly
non-trivial boundary conditions \cite{risken_fokker-planck_1996}.
An other long standing theoretical framework to study irregularly
pulsing units is the theory of the stochastic point processes. In this theory,
the event times are described by probability density functions
which are history dependent. Solutions of the first passage time problem have
long been used to connect this phenomenological description to the underlying
dynamics of the state variables \cite{schrodinger_zur_1915,
*chandrasekhar_dynamical_1943, *wang_theory_1945}.

In this letter, we marry these two approaches, exploiting the fact in that
pulse-coupled systems the recurrent inputs into the units is fully determined by
the timing of events.
The only element that needs to be added to the first passage time description,
is the self-consistency of the interactions and the network output.  We first
demonstrate our method on a simple system and show that the description of
asynchronous state and its stability is consistent with previously derived
results \cite{abbott_asynchronous_1993, brunel_sparsely_2008}. We then add a
dynamic component to the interactions. This addition is not readily incorporated
within the Fokker-Plank approach, however, it is easily incorporated in our
formalism. 
Such dynamic recurrent couplings can be observed in many physical
systems. For instance, temporal dynamics of intracellular signaling activities
is tightly regulated by positive or negative feedback
\cite{kholodenko_cell-signalling_2006}, similarly biochemical processes
concerning transmitter production and release in synapses in the network of
neurons are known to be modulated by the activity of interacting cells
\cite{grossberg_production_1969, tsodyks_neural_1998}.

We consider a network of $N$ identical oscillators with all-to-all feedback 
coupling, which receives a noisy external input. We assume that the 
oscillators are modeled as integrate and fire neurons, where their the membrane
voltage is the state variable. Between events the 
(normalized) voltage $x_i$, of oscillator $i$ satisfies
\begin{equation}
\taum\frac{d}{dt}x_i=\iota_i(t)-x_i,
\end{equation}
where $\taum$ is the membrane time constant and $\iota_i$ is the input 
current into oscillator $i$. When the voltage reaches the threshold, $x_\thr=1$,
the oscillator emits a pulse and the voltage is immediately reset to the resting
potential, $x_\rest=0$\@.
The input, $\iota_i$, can be written as  $\iota_i=\iota_{i,\ext}+\iota_{i,\fb}$
where $\iota_{i,\ext}$ and $\iota_{i,\fb}$ are the external and feedback input
respectively.
The  external current is given by
$\iota_{i,ext}(t)=\mu_\ext+\sigma\eta_i(t)$, where the $\eta_i$s are
independent Gaussian white noise variables, 
$\langle\eta_i(t)\rangle=0$ and $\langle\eta_i(t)\eta_j(t^\prime)\rangle=
\delta_{ij}(t-t^\prime)$.
When a oscillator emits a pulse at time $t_k$, this causes a, so-called 
synaptic, current input $s(t-t_k)$ in all oscillators. 
This input is given by
\begin{equation}
s(t)=\frac{\gs}{N}\frac{1}{\tausone-\taustwo}\left(e^{-t/\tausone}-
e^{-t/\taustwo}\right)\Theta(t),
\end{equation}
where $\Theta$ is the Heaviside function.
Here $\taustwo$ and $\tausone$ are the synaptic rise and decay
times. 
We study the network in the thermodynamic limit ($N\rightarrow \infty$).
In this limit the total recurrent input into all oscillators is identical,
$\iota_{i,\fb}= \mu_\fb$, and is given by
\begin{equation}
\left(1+\tausone\frac{d}{dt}\right)\left(1+\taustwo\frac{d}{dt}\right)
\mu_\fb(t)=\gs r(t),
\end{equation}
where $r(t)=N^{-1}\sum_i\sum_k\delta(t-t_{i,k})$ is the population 
firing rate.
Here, $t_{i,k}$ is the time of the $k$th event of oscillator $i$\@.

To calculate \textit{inter-event density}, we use that when oscillator $i$
emits a pulse at time $\tp$, we have that $x_i(\tp)=0$ and $x_i$ satisfies the
stochastic differential equation
$\taum\frac{d}{dt}x_i=\mu(t)-x_i+\sigma\eta_i(t)$ until $x_i$ reaches $1$.
Averaging over the realizations of the noise the probability
density $\rho(x,t|\tp)$ for $x_i(t)=x$ and no event has occurred
between $\tp$ and $t$ satisfies the Fokker-Planck equation
\begin{equation}
\taum\frac{\partial}{\partial t}\rho(x,t|\tp)=
-\frac{\partial}{\partial x}[\mu(t)-x]\rho(x,t|\tp)+
\frac{\sigma^2}{2}\frac{\partial}{\partial x^2}\rho(x,t|\tp),
\label{FP:eq}
\end{equation}
with initial condition $\rho(x,t^\prime|t^\prime)=\delta(x)$ and boundary  
condition $\rho(1,t|t^\prime)=0$.
The difficulty in solving this is in satisfying the boundary condition.
For an unrestricted process, which satisfies Eqn. (\ref{FP:eq}) it is 
straightforward to show that with initial condition $x_i(\tp)=\xp$
the probability density $\hat{\rho}(x,t|\xp,\tp)$ for
$x_i(t)=x$ satisfies, for $t\geq\tp$
\begin{equation}
\hat{\rho}(x,t|\xp,\tp)=\frac{1}{\sqrt{2\pi}\Sigma(t-\tp)}
\exp\left(\frac{-(x-\xav(t|\xp,\tp))^2}{2\Sigma^2(t-\tp)}
\right),
\label{rhohat:eq}
\end{equation}
where the noise-averaged of $x$ is denoted as $\xav$ and it satisfies  
\begin{equation}
\xav(t|\xp,\tp)= \xp e^{-(t-\tp)/\taum}+\frac{1}{\taum}\int_{\tp}^t\!
d\tpp\,\mu(\tpp)e^{-(t-\tpp)/\taum}
\end{equation} 
and the variance $\Sigma^2$ is given by
\begin{equation}
\Sigma^2(t)= \frac{\sigma^2}{2\taum}
\left[1-e^{-2t/\taum}\right].
\label{sigma:eq}
\end{equation}
The problem with initial condition $\rho(x,\tp|\tp)=\delta(x)$ and an
absorbing boundary at $x=1$ can be viewed as an unrestricted process, where a 
particle is inserted at $x=0$ at time $\tp$ and
extracted at $x=1$ at time $t$ with some probability density ${\Pr}(t|\tp)$ so
that
\begin{equation}
\rho(x,t|\tp)=\hat\rho(x,t|0,\tp)
-\int_{\tp}^t\!d\tpp\, {\Pr}(\tpp|\tp)\hat\rho(x,t|1,\tpp).
\end{equation}
The inter-event probability density ${\Pr}(t|\tp)$ is determined by the
boundary condition, $\rho(1,t|\tp)=0$ and thus satisfies the Volterra
integral equation
\begin{equation}
\hat\rho(1,t|0,\tp)=\int_{\tp}^t\!d\tpp\, \hat\rho(1,t|1,\tpp){\Pr}(\tpp|\tp).
\label{volterra:eq}
\end{equation}

In the \textit{asynchronous state} the emission rate is constant, $r(t)=r_\eq$,
and we have $\mu(t)=\mu_\eq=\mu_\ext+g_s r_\eq$\@. Consequently, $\hat{\rho}$ is
given 
by $\hat\rho(x,t|\xp,\tp)=\hat\rho_\eq(x,t-\tp|\xp)$, where 
$\hat\rho_\eq(x,t|\xp)$ satisfies Eqn. (\ref{rhohat:eq}) with 
$\xav(t|\xp,\tp)=\xaveq(t-\tp|\xp)=\mu_\eq+(\xp-\mu_\eq)\exp(-[t-\tp]/\taum)$\@.
Note that for $t \rightarrow \infty$, 
$\hat\rho_\eq(x,t|\xp)\rightarrow \hat\rho_\infty(x)
\equiv \exp(-(x-\mu_\eq)^2/2\Sigma_\infty^2)/\sqrt{2\pi}\Sigma_\infty$, where
$\Sigma_\infty=\sigma/\sqrt{2\taum}$,
Additionally, the inter-event probability density can be written as
${\Pr}(t|\tp)={\Pr}_\eq(t-\tp)$ as this density is time invariant in the
stationary asynchronous regime. This density must satisfy the following Volterra
integral equation
\begin{equation}
\hat\rho_\eq(1,t|0)=\int_0^t\!d\tp\hat\rho_\eq(1,t-\tp|1){\Pr}_\eq(\tp).
\end{equation}
The right hand side of this equation is now a convolution and this can be 
solved using the Laplace transform. The Laplace transform
${\Pr}_{\eq,\Laplace}$ 
of ${\Pr}_\eq$ satisfies
\begin{equation}
{\Pr}_{\eq,\Laplace}(s)=\frac{\hat\rho_{\eq,\Laplace}(1,s|0)}
{\hat\rho_{\eq,\Laplace}(1,s|1)},
\label{ISIlaplace:eq}
\end{equation}
where $\hat\rho_{\eq,\Laplace}$ is the Laplace transform of $\hat\rho_\eq$.
In the supplementary material \cite{[{Supplementary
matrial at}] supp}, we show that the Laplace transform of Ornstein-Uhlenbeck
density $\hat\rho_{\eq,\Laplace}$, can be formally calculated (for an
alternative derivation refer to \cite{tuckwell_introduction_2005}).
Now, we can close the system, since the average inter-event interval, $\langle
t\rangle$, can be found using  $\langle t\rangle=-{\lim}_{s\rightarrow 0}d
{\Pr}_{\eq,\Laplace}(s)/ds$ and this  allows to express $r_\eq=1/\langle
t\rangle$ in terms of $\mu_\eq$ and $\sigma$\@, together with
$\mu_\eq=\mu_{ext}+g_s r_\eq$ that determines $r_\eq$\@.

To determine the \textit{stability} of the asynchronous state we consider the
evolution of small perturbations around the equilibrium firing rate,
$r(t)=r_\eq+\epsilon
r_1(t)$. 
With such a perturbation the noise averaged input into oscillators
$\mu(t)$ 
satisfies $\mu(t)=\mu_\eq+\epsilon \mu_1(t)$, where $\mu_1$ is given by
\begin{equation}
\mu_1(t)=\gs \frac{1}{\tausone-\taustwo}\int_0^\infty\!d\tp\,
r_1(t-\taud-\tp)(e^{-\tp/\tausone}-e^{-\tp/\taustwo}).
\end{equation}
The probability density $\hat\rho$ for the unconstrained diffusion
still satisfies Eqn. (\ref{rhohat:eq}), but now with
$\bar{x}(t|\xp,\tp)=\bar{x}_\eq(t-\tp|\xp)+ \epsilon\bar{x}_1(t|\tp)$,
where $\bar{x}_1$ satisfies
\begin{equation}
\bar{x}_1(t|\tp)=\int_{\tp}^t\!d\tpp \mu_1(\tpp)e^{-(t-\tpp)/\taum}.
\end{equation}
Thus, we can expand $\hat\rho$ as
$\hat\rho(x,t|\xp,\tp)=\hat\rho_\eq(x,t-\tp|\xp)
+\epsilon\hat\rho_1(x,t|\xp,\tp)+O(\epsilon^2)$, where
\begin{equation}
\hat\rho_1(x,t|\xp,\tp)=-\bar{x}_1(t|\tp)\frac{\partial}{\partial x}\hat\rho
(x,t-\tp|\xp).
\end{equation}
Next, we write for the inter-events probability density
${\Pr}(t|\tp)={\Pr}_\eq(t-\tp)+\epsilon {\Pr}_1(t|\tp) +O(\epsilon^2)$ and
insert
this with the expansion for $\hat\rho$ in Eqn.  (\ref{volterra:eq}),
to obtain for ${\Pr}_1(t|\tp)$ the Volterra integral equation
\begin{eqnarray}
\hat\rho_1(1,t|0,\tp)&-&\int_{\tp}^t\!d\tpp\,\hat\rho_1(1,t|0,
\tpp){\Pr}_\eq(\tpp-\tp)\nonumber \\&=&
\int_{\tp}^t\!d\tpp\,\hat\rho_\eq(1,t-\tpp|0){\Pr}_1(\tpp|\tp).
\label{volterra1:eq}
\end{eqnarray}
Finally, we close the system using $r(t)$, that is
\begin{equation}
r_\eq+\epsilon r_1(t)= \int_{-\infty}^t\!d\tp\,[{\Pr}_\eq(t|\tp)+\epsilon
{\Pr}_1(t|\tp)]
[r_\eq+\epsilon r_1(\tp)]+O(\epsilon^2).
\label{ratepert:eq}
\end{equation}
For sufficiently small $\epsilon$, we can ignore terms of order
$\epsilon^2$.  In this linearized system, we can make the usual
  Ansatz that $r_1(t)=r_\lambda e^{\lambda t}$. With this 
  Ansatz we will, as we will see below, have $\mu_1(t)=r_\lambda
e^{\lambda t}\mu_\lambda$, $\hat\rho_1(x,t|\xp,\tp)=r_\lambda
e^{\lambda t}\hat\rho_\lambda(x,t-\tp|\xp)$ and ${\Pr}_1(t|\tp)=r_\lambda
e^{\lambda t} {\Pr}_\lambda(t-\tp)$. Inserting this in
Eqn. (\ref{ratepert:eq}) we obtain the eigenvalue equation
\begin{equation}
\int_0^\infty\!dt\,{\Pr}_\eq(t)e^{-\lambda t}+r_\eq \int_0^\infty\!dt\,
{\Pr}_\lambda(t)=1.
\label{eigen:eq}
\end{equation}
With $r_1(t)=r_\lambda e^{-\lambda t}$ we have
\begin{equation}
\mu_1=r_\lambda e^{-\lambda t}\frac{\gs e^{-\lambda\taud}}{(1+\lambda\tausone)
(1+\lambda\taustwo)}
\end{equation}
and
\begin{equation}
\bar{x}_1(t|\xp,\tp)=r_\lambda e^{-\lambda t} A_\lambda(t, \tp),
\end{equation}
where $A_\lambda(t,\tp)=\frac{\gs e^{-\lambda\taud}
  (1-e^{-(\lambda+1/\taum)(t-\tp)})}
{(1+\lambda\tausone)(1+\lambda\taustwo)(1+\lambda\taum)}$.
Thus, we can write $\hat\rho_1$ as $\hat\rho_1(x,t|\xp,\tp)=r_\lambda
e^{-\lambda t} \hat\rho_\lambda(x,t-\tp|\xp)$ with
\begin{equation}
\hat\rho_\lambda(x,t-\tp|\xp)=-A_\lambda(t, \tp)\frac{\partial}
{\partial x} \hat\rho_{\eq}(x,t-\tp|\xp)
\end{equation}
We insert this into Eqn. (\ref{volterra1:eq}), we find hat ${\Pr}_1$ is
given by ${\Pr}_1(t|\tp)=r_\lambda e^{-\lambda t}{\Pr}_\lambda(t-\tp)$, where
${\Pr}_\lambda$ satisfies
\begin{eqnarray}
  \int_0^t\!d\tp\,\hat\rho_\eq(1,t-\tp|1)e^{-\lambda(t-\tp)}{\Pr}_\lambda(\tp)
\nonumber \\=
  \hat\rho_\lambda(1,t|0)-\int_0^t\!d\tp\,\hat\rho_\lambda(1,t-\tp|1)
  {\Pr}_\eq(\tp).
\end{eqnarray}
Multiplying both sides by $e^{-st}$ and integrating over $t$ we find for
the Laplace transform ${\Pr}_{\lambda\Laplace}$ of ${\Pr}_\lambda$
\begin{equation}
{\Pr}_{\lambda,\Laplace}(s)=\frac
{\hat\rho_{\lambda,\Laplace}(1,s|0)-\hat\rho_{\lambda,\Laplace}(1,s|1)
{\Pr}_{\eq,\Laplace}(s)}{\hat\rho_{\eq,\Laplace}(1,s+\lambda|1)}.
\end{equation}
Here $\hat\rho_{\lambda,\Laplace}$, the Laplace transform of $\hat\rho_\lambda$
that satisfies
\begin{equation}
\hat\rho_{\lambda,\Laplace}(x,s|\xp)=-\frac{\partial}{\partial x} B_\lambda(
\hat\rho_{0,\Laplace}(x,s|\xp)-
\hat\rho_{0,\Laplace}(x,s+\lambda+\taum^{-1}|\xp))
\end{equation}
where $B_\lambda=\frac{\gs
  e^{-\lambda\taud}}{(1+\lambda\tausone)(1+\lambda\taustwo)(1+\lambda\taum)}$.
We can rewrite the eigenvalue equation (Eqn. \ref{eigen:eq}) as
${\Pr}_{0,\Laplace}(\lambda)+
r_\eq {\Pr}_{\lambda,\Laplace}(0)=1$, and plugging it in the expressions for
the Laplace transforms obtained above, we find that the straightforward
eigenvalues,
$\lambda$s, of the system that satisfy
\begin{eqnarray}
 \gs r_\eq \left. \int_0^\infty\! dt\,\kappa_1 D_x(t) = 
\kappa_2 
 \int_0^\infty\! dt\, e^{-\lambda t}D_1(t),\right.
\label{eigen:eq2}\end{eqnarray}
where $\kappa_1 =1-e^{-(\lambda+\taum^{-1})t}$, $\kappa_2=
e^{\lambda\taud}(1+\lambda\tausone)(1+\lambda\taustwo)(1+\lambda\taum)$,
$D_x(t) = \frac{\partial}{\partial x}\left.(\hat\rho_\eq(x,t|0)-
\hat\rho_\eq(x,t|1)\right|_{x=1}$ and 
$D_1(t) = \hat\rho_\eq(1,t|0)-\hat\rho_\eq(1,t|1)$. This expression is exact and
it directly corresponds to the eigenvalue equation that has been formally
derived using the perturbation of Fokker-Plank
operator~\cite{abbott_asynchronous_1993}.

Unlike the latter, our approach is easily extended to networks in which the 
recurrent connections are mediated through \textit{coupling with dynamic
strength}.  
In this part we will demonstrate that the stability analysis of these systems
can be treated effortlessly. Here, without loss of generality, we focus on the
couplings with depressive interactions (e.g. negative feedback).
The network is as before, except that the recurrent input due to a event of
oscillator $i$ depends on the strength factor $p_i$, denoted as release factor.
For instance, in network of neurons with depressive couplings the biophysical
meaning of $p_i$ is the amount of vesicles that are available in synapses. The
recurrent input is given by
\begin{equation}
\left(1+\tausone\frac{d}{dt}\right)\left(1+\taustwo\frac{d}{dt}\right)
\mu_\fb(t)=g_s r_r(t)
\end{equation}
where $r_r(t)$ is the rate of release rather then the event
emission rate. The release rate is given by $r_r(t)=u N^{-1}\sum_{i,k}
p_i(t^-)\delta(t-t_{ik})$, where $u$ is the 
release fraction (see \cite{tsodyks_neural_1998}). Between events the
vesicles are replenished with a time constant $\tau_D$,
$dp_i/dt=(1-p_i)\tau_D$ and at the time of the 
event an amount $up_i$ of vesicles is released and $p_i$ is reset to 
$(1-u)p_i$.
It is straightforward to show that if the event density is $\Pr(t|\tp)$, the 
release rate $r_r$ satisfies
\begin{eqnarray}
  r_r(t) &=& u \int d\tp \Pr(t|\tp)(1- e^{-\frac{t-\tp}{\tau_D}})r(\tp) 
 \nonumber \\
  & & \mbox{} + (1-u) \int d\tp \Pr(t|\tp) e^{-(\frac{t-\tp}{\tau_D})}r_r(\tp).
 \label{rp:eq}
\end{eqnarray}
The rate in the asynchronous solution can be determined as follows: Given 
$\mu_\eq$ and $\sigma$ we calculate the Laplace transform $\Pr_\Laplace$ of the 
inter-event distribution as before. This determines the equilibrium rate.
Using this and Eqn. (\ref{rp:eq}), then we obtain for the steady state release
rate 
$r_{r,\eq}=u r_\eq/[1-(1-u)\Pr_\Laplace(1/\tau_D)]$. The self-consistency 
requirement is that this should agree with $\mu_\eq=\mu_\ext+g_s r_{r.\eq}$\@. 
Now, it is also straightforward to extend the stability analysis to this model.
We starts with the Ansatz that the release rate satisfies 
$r_r(t) =
r_{r,eq} + \epsilon r_{r,\lambda}e^{\lambda t}$\@. Following the steps of the 
model with static couplings, we obtain $\Pr(t|\tp)=\Pr_\eq(t-\tp)+
\epsilon e^{\lambda t}\Pr_\lambda(t|\tp)$ and $r(t)=r_\eq+
\epsilon e^{\lambda t} r_\lambda$, with $\Pr_\lambda$ and $r_\lambda$ 
proportional to $r_{r,\lambda}$. Combining this with Eqn. (\ref{rp:eq}) and 
requiring self-consistency leads to the following eigenvalue equation
\begin{eqnarray*}
\lefteqn{\gs r_\eq \left( u \int_0^\infty\! dt\,\kappa_1 \kappa_3 
D_x(t) + (1-u) \int_0^\infty\! dt\,\kappa_1 \kappa_4 D_x(t) \right) = }
\nonumber \\
  & & \kappa_2 \left( u \int_0^\infty\! dt\, \kappa_3 e^{-\lambda t}D_1(t)
+(1-u) \int_0^\infty\! dt\, \kappa_4 e^{-\lambda t}D_1(t) \right),
\end{eqnarray*}
where, $\kappa_3 = 1 - e^{-t/\tau_D}$ and $\kappa_4 = e^{-t/\tau_D}$. Now, we
can assess the stability of the network with dynamic couplings, by finding
$\lambda$s that satisfy the above equation. 
We numerically determined the eigenvalues for different $\tau_D$ and $u$, 
adjusting $\gs$ to jeep the rate constant 
in the asynchronous state. The asynchronous state destabilizes through 
a pair of purely imaginary $\lambda$s,  corresponding to the emergence of s
limit cycle oscillations due to an Andronov-Hopf bifurcation. Figure 1 shows 
the resulting phase diagram.

The approach introduced here can be utilized to evolve a system for 
$N\rightarrow \infty$, as the time dependent inter-event
density can be self-consistently determined. By exploiting this, we
numerically evolve a network with depressive coupling \cite{[{Supplementary
matrial at}] supp} to study the behavior near the bifurcation point. 
The activity continuously changes after the
bifurcation point (Fig.2) and the amplitude of collective synchrony grows 
indicting a super-critical Hopf bifurcation~\cite{[{Supplimanty
matrial at}] supp}.

\begin{figure}
\centering
\includegraphics[scale=.95]{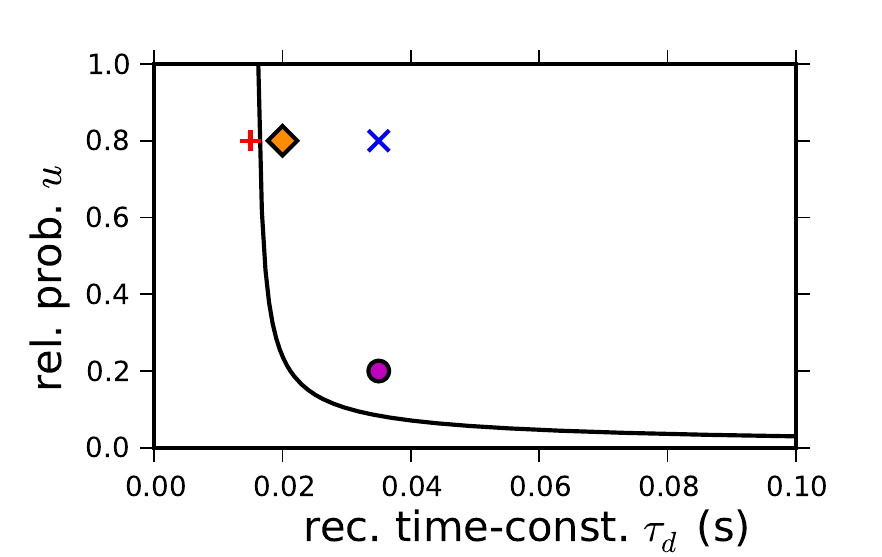}
\caption{\label{fig:phase} Phase diagram of a system with depressive
   couplings. The asynchronous stationary state is stable underneath
   the curve. The solid line correspond to the parameter regime where the
   purely imagery eigenvalues give rise to Hopf bifurcation of asynchronous
irregular state.
   The marked symbols are the parameters in the phase space that we adopt
   to numerically evolve the system \cite{[{Supplimanty
matrial at}] supp} in Fig.2.
   Parameters: $\mu_{ext}$=0.95, $\sigma$=0.0228,
   $\gs$=0.00245, $\tau_m$=0.020 and $\tau_{s2}$=0.001,
   $x_\thr$=1 and $V_r$=0.
}
\end{figure}

\begin{figure}
\centering
\includegraphics[scale=.95]{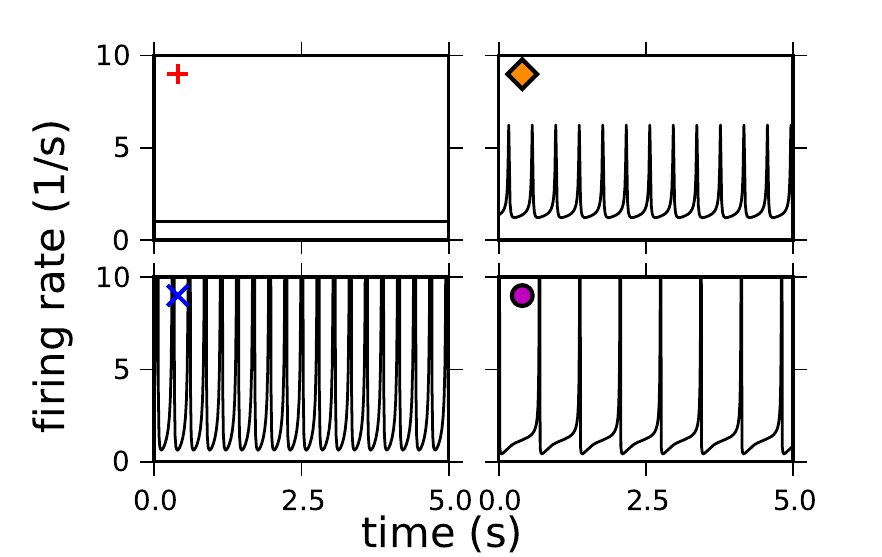}
\caption{\label{fig:pop_sim} The numerical
   simulation of the full system with depressive couplings using population
density treatment \cite{[{Supplimanty
matrial at}] supp}. Each subplot
   illustrates the population firing rate of the system with the parameters are
marked with corresponding symbol in Fig.1.}
\end{figure}

In the present letter, we derived self-consistent description of the inter-event
distribution of non-linear pulse coupled oscillators with interactions. We
additionally characterized the asynchronous state and its stability. For 
static interactions, this result coincides with the result from the classical
Fokker-Planck approach. However, as we
showed our approach is easily extended to incorporate the effect of dynamical
coupling. Using this, we investigated how networks with interaction through
couplings with short-term depression undergoes a Hopf bifurcation to a state
with collective synchronization, so-called population spikes.
The method also allows for a efficient way to simulate
the dynamics of systems in their thermodynamic limits~\cite{[{Supplimanty
matrial at}] supp}. We showed in this limit the system may exhibit a
super-critical Hopf bifurcation.
Up to now, the collective effect of activity dependent modulation of interaction
in a network has only been analyzed in models without non-linear contributions
of the inter-event density \cite{cortes_short-term_2013, tsodyks_neural_1998}.
In such networks interesting phenomena such a  as Shilnikov chaos has been 
observed when positive feedback (e.g. facilitation) are added
\cite{cortes_short-term_2013}. 
It is straightforward to extend our approach to study those cases where the
network of oscillators with self-consistent inter-event density (e.g. spiking
neurons) are also considered.
The method presented in here relies on our ability to calculate
$\hat\rho$, the solution of the unrestricted of Fokker-Planck equation. Once 
this is achieved imposing the necessary boundary conditions is easy using
the presented method. Thus we believe that our approach will have a wide range 
of applications.

\textit{Acknowledgment}. FF was funded by GIF-I-1224-396.13/2012. CvV
was supported by ANR-BALWM grant.

\bibliography{novel_ref}

\end{document}